%% file: arXiv-CSQCSII.tex
\documentclass[12pt]{article}
\usepackage{epsfig}

\newcommand{\be}{\begin{eqnarray}}
\newcommand{\ee}{\end{eqnarray}}

\input econfmacros.tex
\def\Title#1{\begin{center} {\Large {\bf #1} } \end{center}}

\begin{document}

\Title{Lattice QCD and dense quark matter}

\bigskip\bigskip


\begin{raggedright}

{\it M.P. Lombardo\index{Lombardo, M.P.}\\
INFN-Laboratori Nazionali de Frascati\\
I-00044, Frascati (RM)\\
Italy}
\bigskip\bigskip
\end{raggedright}

\begin{raggedright}

{\it K. Splittorff (speaker) \index{Splittorff, K.}\\
The Niels Bohr Institute\\
University of Copenhagen\\
Blegdamsvej 17 \\
DK-2100, Copenhagen \\
Denmark\\
}
\bigskip\bigskip
\end{raggedright}

\begin{raggedright}

{\it J.J.M. Verbaarschot\index{Verbaarschot, J.J.M.}\\
Department of Physics and Astronomy\\
SUNY, Stony Brook\\
New York 11794\\
USA\\
}
\bigskip\bigskip
\end{raggedright}

\abstract{This talk summarizes recent progress in lattice QCD for dense
  quark matter. The emphasis is on the insights obtained from analytical 
  results derived within chiral perturbation theory.}

\section{Introduction}

Despite the fact that all observed processes conserve baryon number,
the total number of baryons present today is far greater than the number of
anti-baryons. It is a long standing challenge to explain this excess of
baryons. If we accept the baryon imbalance as a fact and ask:
\vspace{2mm}

{\sl What is the preferred ground state of strongly interacting
matter given a specific baryon density and temperature ?} 
\vspace{2mm}

\noindent
we are faced
with an equally long standing challenge: The phases of strongly
interacting matter must 
be determined by non-perturbative means, and the only known first
principle method, lattice QCD, is extremely challenging to apply 
when there is an imbalance between baryons and anti-baryons. 

Even though these two challenges are closely related their solutions
will presumable be of a very different nature. To understand the 
generation of the baryon number most likely requires physics beyond
the standard model. To understand the implications  
of a non zero baryon number on the ground state of strongly
interacting matter requires the development of new non-perturbative
tools for QCD. 

This talk focuses on the latter challenge so let us explain in detail what we
are up against when we seek to describe baryon asymmetric matter from 
first principles (even when we are modest, in that we do not aim to 
explain where this asymmetry originally came from, but simply want to
adopt this into QCD). For simplicity we will also ignore electromagnetism.  

Rather than fixing the density the natural approach is to start with
the Grand Canonical partition function, Z, where we fix the quark
chemical potential, $\mu$, and then determine the quark number $n$ 
(the baryon number is 3$n$). The average quark number is  
\begin{equation}
\langle n \rangle  = \partial_\mu \log Z(\mu).
\end{equation}
It is straightforward to include $\mu$ in the partition
function. Since the chemical potential is conjugate to the 
quark number, the combination that appears in the Lagrangian is 
\begin{equation}
\mu n = \mu q^\dagger q  = \mu  \bar{q}\gamma_0 q. 
\end{equation}
The simple extension of the Dirac operator in the Lagrangian of QCD 
\begin{equation}
{}{\cal L}_{\rm QCD} 
= \bar{q}(D_\eta\gamma_\eta+\mu\gamma_0+m)q + {\rm Gluons} 
\end{equation}
thus accommodates the chemical potential into QCD.

It is well known how to implement this extension on the lattice
\cite{HK,KMSWSSS}. Since $\mu$ enters as the zeroth component of 
a vector potential the choice with the correct continuum limit
is \cite{HK}  
\begin{equation}
{}{\cal L}_{\rm Lattice \hspace{1mm} QCD} 
= ... + e^{a\mu}\bar{q}_x\gamma_0U_{x,x+\hat{0}}q_{x+\hat{0}}
     + e^{-a\mu}\bar{q}_{x+\hat{0}}\gamma_0U^\dagger_{x,x+\hat{0}}q_x 
+ ...  
\end{equation}
For two color QCD, when the fermion determinant is real, lattice
simulations \cite{HKLM,KTSI,KTSII,HandsI,HandsII} based on this  
discretization give a rich phase diagram that agrees with theoretical
expectations \cite{KST,KSTVZ,LSS}. For three colors, however, the fermion
determinant at non zero chemical potential is not real and positive   
\begin{equation}
{\det}^2(D+\mu\gamma_0+m) =  |{\det}(D+\mu\gamma_0+m)|^2 e^{2i\theta}.
\end{equation}
While there is nothing wrong physically with a complex valued fermion
determinant it is a major problem for lattice QCD. 
The Monte Carlo sampling of gauge field configurations, which is the
back bone of lattice QCD, only works if the measure 
in the Euclidean partition function is real and positive. This is
{\sl the QCD sign problem}: at $\mu\neq0$ direct Monte Carlo sampling
is not possible and standard lattice QCD breaks down.

Several methods have been engineered to circumvent the sign problem,
see table \ref{tab:methods}. Each have given important first principle
insights into QCD at non zero $\mu$, and a consistent picture emerges at
small $\mu/T$. All methods face serious challenges more or less
directly induced by the sign problem. In order to deal with these
challenges analytical results are of utmost importance. For example, in the
method of Ejiri \cite{Ejiri} the distribution of the phase of the
fermion determinant is assumed to be Gaussian. Here we show that the
Gaussian form comes out analytically for small chemical potentials,
thus confirming the assumptions of this lattice method. However, as we
also show, the distribution changes to a Lorentzian form for larger
$\mu$. This computation is carried out within chiral perturbation
theory, which 
is the low energy effective theory for QCD in the chirally broken
phase. The analytic results in this way both justifies the assumptions
on which the method of Ejiri is based and show its limitations. 

Chiral perturbation theory also allows us to understand the
distribution of the values which the Euclidean quark number operator assumes
over the gauge fields. The results not only give us a direct insight in
the way the total baryon number forms, it also shows how the Complex
Langevin method can deal with the sign problem.

\begin{table}[t]
\begin{tabular}{|c|c|c|}
\hline
&& \\
{\large \bf Method} & {\large \bf Idea} & {\large \bf Challenge} \\
&& \\
\hline
&& \\
{Reweighting} & {\small Absorb the sign in the observable} &
{\small Exponential cancellations} 
 \\
\cite{Glasgow,fodor1,fodor2} && \\
\hline
&& \\
{Taylor expansion} & {\small Expand at $\mu=0$} & {\small Higher order terms}
   \\
\cite{gupta,Allton1,Allton2,Allton3,Endrodi:2009sd} && \\
\hline
&& \\
{Imaginary $\mu$} & {\small Determine the analytic function} &
 {\small Control the extrapolation} \\
\cite{owe1,maria,owe2,ERL}&& \\
\hline
&& \\
{Density of states} & {\small Use the distribution of the phase} & 
{\small Determine the distribution} \\
\cite{Gocksch,Azcoiti,AN,AANV,Schmidt,Ejiri}&& \\
\hline
&& \\
{Canonical ensemble} & {\small Work at fixed baryon number} &  {\small
  Fix the baryon number}\\   
\cite{MR,HT,EKKL,KdeF,LMAL} &&  \\
\hline
&& \\
{Complex Langevin} & {\small Stochastic flow in complex plane} &
{\small Make it work for QCD} \\ 
\cite{KW,FOC,AFP,aarts} && \\
\hline
\end{tabular}
\caption{\label{tab:methods}Summary of the main methods used to
  circumvent the sign problem, the main idea used and the main
  challenge faced when using this method.}
\end{table}

\section{Fixed phase of the determinant}

Perhaps the most direct way to understand the severity of the QCD sign
problem is to consider the distribution of the phase of the fermion
determinant 
\begin{equation}\label{rhoTh1} 
\langle\delta(\theta-\theta')\rangle_{1+1}d\theta
=\frac{\int dA |\det(D+\mu\gamma_0+m)|^{2} e^{2i\theta'}
  \delta(\theta-\theta') e^{-S_{\rm YM}}}
{\int dA |\det(D+\mu\gamma_0+m)|^{2} e^{2i\theta'} e^{-S_{\rm
      YM}}} d\theta . 
\end{equation}
The $\delta$ function allows us to rewrite the unquenched distribution
in terms of the phase quenched one times a phase factor and a normalization
\begin{equation}\label{th-distBasic}
 \langle \delta(\theta-\theta')\rangle_{1+1} = e^{2i\theta
 }\frac{Z_{1+1^*}}{Z_{1+1}}\langle \delta(\theta-\theta')\rangle_{1+1^*}.  
\end{equation}
The complex nature of the unquenched $\theta$ distribution is typical
of the sign problem. Here it takes the simplest possible
form. Nevertheless, the effect of the phase is dramatic: it leads to
exponentially large cancellations in the volume. This follows directly
since $Z_{1+1^*}/Z_{1+1}\sim\exp(V)$. These dramatic cancellations 
are what makes the sign problem severe.

To understand how the cancellations take part we have computed the  
$\theta$ distributions to leading order in chiral perturbation
theory. 
For $\mu<m_\pi/2$ with the phase angle $\theta\in[-\pi,\pi]$ we have 
\cite{1loop}
\begin{eqnarray}
\langle\delta(\theta-\theta')\rangle_{1+1} 
&=& \frac 1{\sqrt{\pi \Delta G_0}} e^{2i\theta+\Delta G_0}
\sum_{n=-\infty}^\infty e^{-(\theta+2n\pi )^2/\Delta G_0},
\label{rhoth-CPT}
\end{eqnarray}
where
\begin{eqnarray}
\Delta G_0
& = & \frac{Vm_\pi^2T^2}{\pi^2}\sum_{n=1}^\infty 
\frac{K_2(\frac{m_\pi n}{T})}{n^2}(\cosh(\frac{2\mu n}{T})-1).   
\label{g0familiar}
\end{eqnarray}
Notice that $\Delta G_0$ is the difference of the free energy in the
full and the phase quenched theory
\begin{equation}
\frac{Z_{1+1}} {Z_{1+1^*}} = e^{-\Delta G_0}.
\end{equation}
The $\theta$ distribution takes the form of a Gaussian modulo
$2\pi$. This is in agreement with observations on
the lattice \cite{Ejiri} and is the natural form suggested by the 
central limit theorem. However, at larger values of the chemical
potential the quark mass enters the support of the Dirac spectrum  
\cite{Gibbs,TV} and fluctuations of the phase will be far greater
\cite{PhaseLetter,PhaseLong,LSVI,LSVII}.

To leading order in chiral perturbation theory for $\mu>m_\pi/2$ the
$\theta$ distribution is a Lorentzian modulo 2$\pi$ times the phase
factor \cite{LSVI}:  
\begin{equation}\label{th-distmuGmpi2UQ}
\langle\delta(2\theta-2\theta')\rangle_{1+1}
= e^{2i\theta}
\frac{e^{V L_B}}{\pi} \frac{\sinh(V L_B)}{\cosh(V L_B)-\cos(2\theta)},
\end{equation}
where
\begin{equation}
\frac{Z_{1+1}} {Z_{1+1^*}} = e^{-V L_B}.
\end{equation}
This demonstrates that the assumptions of the central limit theorem
are not satisfied and that the method of Ejiri must be examined
carefully for $\mu>m_\pi/2$. 

\begin{figure}[t]
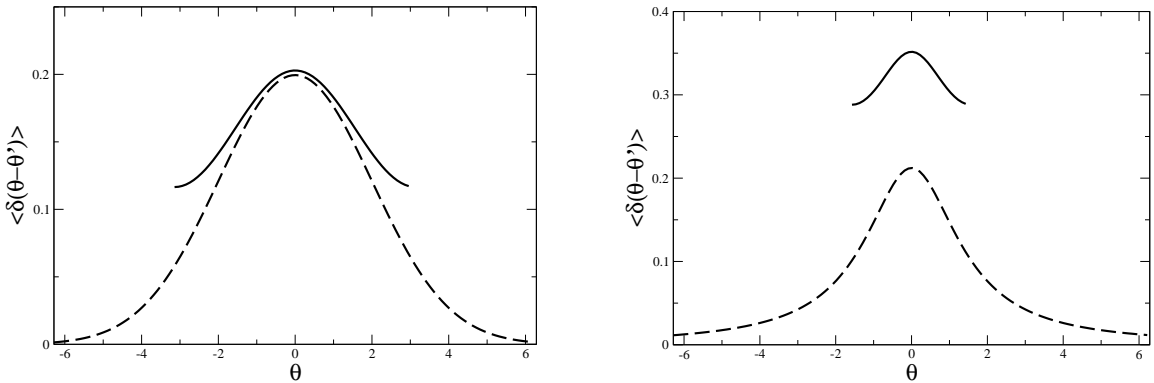

\begin{center}
\epsfig{file=gaussian-comp.eps,height=5cm} \hfill 
\epsfig{file=lorentz-comp.eps,height=5cm}  
\caption{The quenched distribution of the phase of the fermion
  determinant. The dashed curves are standard Gaussian (left) and
  Lorentzian (right). The full curves are the corresponding
  distributions modulo 2$\pi$ and $\pi$ relevant for $\mu<m_\pi/2$ and 
$\mu>m_\pi/2$ respectively.}
\label{fig:theta-dist}
\end{center}
\end{figure}

The Gaussian form (for $\mu<m_\pi/2$) and the Lorentzian form (for
$\mu<m_\pi/2$) modulo $2\pi$ are illustrated in Fig.~\ref{fig:theta-dist}.


\section{Distribution of the quark number over $\theta$}

Since chiral perturbation theory deals with pions alone, the quark
number vanishes (we do not consider the effect of pion fields with
non-trivial topological winding). However, the distribution of the quark
number over $\theta$ is non trivial even when evaluated within chiral
perturbation theory. The form it takes gives us direct insight into 
the noise produced by the pions in lattice QCD.

With $\mu<m_\pi/2$ we find to leading order in chiral perturbation
theory \cite{LSVI}
\begin{equation}
\langle n \ \delta(\theta-\theta')\rangle_{1+1} 
 = \nu_I \sum_{n=-\infty}^{\infty}(1+i\frac{\theta+2\pi
   n}{\Delta G_0})\frac{e^{2i\theta}}{\sqrt{\pi \Delta
    G_0}}e^{-(\theta+2\pi n)^2/\Delta G_0+\Delta G_0},
\end{equation}
where 
\begin{equation}
\nu_I = \left(\lim_{\tilde{\mu}\to\mu} \frac{d}{d\tilde{\mu}} \Delta
   G_0(-\mu,\tilde{\mu})\right).
\end{equation}
This distribution gives the quark number measured on a set of
configurations for which the phase of the determinant is between
$\theta$ and $\theta+d\theta$. 

Note that the normalization is by the two flavor partition function,
so that the total quark number is the integral over the phase angle.

For $\mu>m_\pi/2$ the distribution again changes its form drastically
\cite{LSVI} 
\begin{eqnarray}
&&\langle n \ \delta(2\theta-2\theta')\rangle_{1+1} 
=-\frac{2}{\pi} \Big[\frac{VL_I}{\mu}\Big]
e^{2i\theta}e^{2VL_B} \frac{1}{e^{VL_B}-e^{2i\theta}}. 
\end{eqnarray}

Prior to discussing the importance of these results let us present the
quark number distribution as predicted by chiral perturbation theory.  

\section{Distribution of the quark number}

The distribution of the values which the quark number operator,     
\begin{equation}\label{ndef}
n(\mu)\equiv {\rm Tr}\,\frac{\gamma_0}{D+\mu\gamma_0+m},
\end{equation}
assumes as the gauge fields vary is also accessible within chiral
perturbation theory.

Since 
\begin{equation}\label{nBherm}
n(\mu)^*=\left({\rm Tr}\frac{\gamma_0}{D+\mu\gamma_0+m}\right)^*
={\rm Tr}\frac{\gamma_0}{D-\mu\gamma_0+m}=-n(-\mu),
\end{equation}
the quark number operator is in general complex (it is purely
imaginary at $\mu=0$). 
The fluctuations of the quark number thus occur in the complex
plane. The distribution of the quark number in the complex plane is by 
definition 
\begin{equation}
P^{1+1}_{n}(x,y)\equiv\left\langle
\delta\left(x-\frac{1}{2}(n(\mu)-n(-\mu))\right)\delta\left(y+i\frac{1}{2}(n(\mu)+n(-\mu))\right)\right\rangle_{1+1}.
\end{equation}
In order to evaluate this average within chiral perturbation theory both
$\delta$ functions are expanded in terms of the moments of the real
and imaginary parts of the quark number. The first moment,
i.e.~the average quark number, vanishes. The higher moments are
off-diagonal susceptibilities and are non-trivial even within chiral
perturbation theory \cite{LSVII}.

For $\mu<m_\pi/2$ we find to 1-loop order \cite{LSVII}
\begin{eqnarray}
\label{PnB}
P^{1+1}_{n}(x,y)=\frac{1}{\pi\sqrt{(\chi_{ud}^I)^2-(\chi_{ud}^B)^2}}
e^{-(x-\nu_I)^2/(\chi_{ud}^I+\chi_{ud}^B)}e^{(iy+\nu_I)^2/(\chi_{ud}^I-\chi_{ud}^B)},
\end{eqnarray}
where
\begin{eqnarray} 
\label{notation-n}
\chi^B_{ud} &\equiv& \left .
 \frac {d^2} {d\mu_1d\mu_2} \Delta G_0(\mu_1, \mu_2)
 \right |_{\mu_1 = \mu_2 =\mu},\\
\chi^I_{ud} &\equiv& \left . 
 \frac {d^2} {d\mu_1d\mu_2} \Delta G_0(-\mu_1, \mu_2)
 \right |_{\mu_1 = \mu_2 =\mu}.
\end{eqnarray}
Note that the distribution factorizes into the  distribution of the
real and imaginary part of the quark number.

For $\mu>m_\pi/2$ the higher moments diverge and lead to power law
tails of the distribution of the quark number operator \cite{LSVII}. 
In the quenched case, for example, the distribution has a inverse
cubic tail.

In the confined phase the pions dominate the
free energy and the quark number is thus generated by subleading
terms. A central aspect of the sign problem is to pick up this small
baryon signal from the background produced by the pions. The results
presented above give the analytical form of the pion background and thus the 
foundation for extracting the quark number.         

\section{Cancellations}

The non trivial results for the distribution of the quark number with
$\theta$ and the distribution of the quark number itself should of
course be consistent with zero average quark number. Therefore, upon
integration over $\theta$ and over the complex quark number plane
respectively, we must find a vanishing average quark number. Here we show
analytically how these cancellations occur.

\subsection{Integration over $\theta$}

The total quark number is obtained from its distribution over $\theta$
by integration over the phase angle
\begin{eqnarray}
\langle n \rangle_{1+1} = \int d\theta \langle n \
\delta(\theta-\theta')\rangle_{1+1}. 
\end{eqnarray}
We now discuss how this becomes zero within chiral perturbation
theory.

First we consider the case $\mu<m_\pi/2$. Here we can write the
distribution of the quark number over $\theta$ as a total deriative of
the $\theta$ distribution times $\nu_I$ 
\begin{eqnarray}
\langle n \ \delta(\theta-\theta')\rangle_{1+1} 
  & = &  \nu_I \sum_{n=-\infty}^{\infty}\frac{1}{2i}\frac{d}{d\theta}\frac{e^{2i\theta}}{\sqrt{\pi \Delta
    G_0}}e^{-(\theta+2\pi n)^2/\Delta G_0+\Delta G_0} \\
  & = &  \nu_I\frac{1}{2i}\frac{d}{d\theta}\langle \delta(\theta-\theta')\rangle_{1+1}.
\end{eqnarray} 
From this we see that all phase angles, $\theta$, are essential in
order to obtain the desired vanishing value. Moreover, we see that these exact
cancellations damp the exponentially large amplitude of the
distribution completely (recall that $\Delta G_0$ is extensive). Therefore,
unless the $\theta$ integration is carried out exactly one will find an
exponentially large contribution to the quark number from the pions.

Also for $\mu>m_\pi/2$ we can write the distribution of the baryon
number over $\theta$ as a total derivative. However, this time it
is {\em not} the derivative of the $\theta$-distribution but rather 
\begin{equation}
\langle n \ \delta(2\theta-2\theta')\rangle_{1+1} =\frac{1}{\pi}
\Big[\frac{VL_I}{\mu}\Big] 
e^{2VL_B}\frac {1}{i} \frac d{d\theta}\log( e^{VL_B}-e^{2i\theta}).
\end{equation}
Again this implies that all angles contribute in an essential way to
the full quark number. The need for an exact integration over
$\theta$ is even more urgent for $\mu>m_\pi/2$, since the amplitude now
grows exponentially fast with the volume even at mean field level.    

\subsection{Integration over the complex $n$ plane}

Given the distribution (\ref{PnB}) of the baryon number, $P^{1+1}_{n}(x,y)$,
we obtain the average quark number upon integration over the
complex quark number plane  
\begin{eqnarray}
\langle n \rangle_{1+1} = \int dx dy \ (x+iy) P^{1+1}_{n}(x,y). 
\end{eqnarray}
The analytical result for the distribution shows precisely how the 
vanishing expectation value of the quark number is realized
\begin{equation}\label{vevnB1p1}
\langle n\rangle_{1+1} 
= \int dx \ x P^{1+1}_{{\rm Re}[n]}(x) + i \int dy \ y
P^{1+1}_{{\rm Im}[n]}(y) 
= \nu_I + i i\nu_I = 0. 
\end{equation}
We see that the total quark number is obtained only after a detailed
cancellation between the contribution from the real part and the
imaginary part. The challenge for unquenched lattice QCD is to
identify the baryon signal within these massive cancellations due to the
pions. 
The analytical result from chiral perturbation theory shows precisely
how this may be realized within the framework of the Complex Langevin
method, for a detailed discussion see \cite{LSVII}.

\section{Summary}

The description of the macroscopic phases of strongly interacting
matter based directly on the microscopic QCD dynamics is hampered by
the QCD sign problem. 

Despite the complex nature of the sign problem substantial progress has 
been made recently. We have derived the distribution of the
complex phase of the Dirac operator analytically and the distribution
of the quark 
number over the phase, as well as the distribution of the quark number
itself. These results give direct analytical insights into the numerical
challenges faced by the density of states method and the Complex
Langevin method when applied to QCD at non zero chemical potential.        

The new results form a novel branch of applications of
chiral perturbation theory. This branch has its root in the evaluation of
partially quenched averages by the replica method \cite{replica}.

\bigskip
KS would like to thank the organizers of CSQCDII for an inspiring
meeting and the Kavli Institute for Astronomy and Astrophysics at
Peking University for its hospitality during the workshop.

\end{document}

%% file: econfmacros.tex
\def\beq{\begin{equation}}
\def\eeq#1{\label{#1}\end{equation}}
\def\eeqn{\end{equation}}

\def\beqa{\begin{eqnarray}}
\def\eeqa#1{\label{#1}\end{eqnarray}}
\def\eeqan{\end{eqnarray}}

\let\bar=\overbar

\def\Dslash{\not{\hbox{\kern-4pt $D$}}}
\def\dslash{\not{\hbox{\kern-2pt $\del$}}}

\def\ee{e^+e^-}

\def\msb{{\bar{\ssstyle M \kern -1pt S}}}

\usepackage{fancyhdr,graphicx}